%% file: main.tex
\documentclass[conference,letterpaper]{IEEEtran}
\usepackage[T1]{fontenc}
\usepackage{amsmath,amssymb}
\usepackage{graphicx,booktabs,array}
\usepackage{cite}
\usepackage{balance}
\usepackage[hidelinks]{hyperref}
\hypersetup{pdfauthor={},pdftitle={Read Boundaries in Agent Memory: Consistency, Completeness, and Repair},pdfsubject={Whole-recall protection and incremental failure recovery}}

\begin{document}
\raggedbottom
\title{Read Boundaries in Agent Memory:\\Consistency, Completeness, and Repair}
\author{\IEEEauthorblockN{Haoting Qiu}
\IEEEauthorblockA{\textit{Independent Researcher}\\United States\\qiu.haot@northeastern.edu}
\and
\IEEEauthorblockN{Qibai Chen}
\IEEEauthorblockA{\textit{Independent Researcher}\\United States\\qibaic@alumni.cmu.edu}}
\maketitle
\begin{abstract}
Agent-memory services assemble evidence through several reads across connections or stores. An update between those reads can pair an assertion with source content from another revision. We protect the complete recall, including enrichment and materialization, and evaluate the resulting evidence and service behavior in Hindsight and Cognee. A 600-run Cognee study compares unprotected recall, a repeated recall, a read/write lock, and optimistic epoch validation under default incremental updates and full rebuilds. At approximately 20 recalls/s, unprotected default updates yield 199 conflicting responses among 12,010 scheduled recalls with spaced writes and 284 among 12,010 with bursts. The lock and epoch validation each deliver complete, compatible evidence for all 24,020 default-update recalls. With default bursts, epoch validation reduces update p95 by 109 ms while adding 290 ms to recall p95 relative to the lock. A separate 160-run incremental-failure study distinguishes mixed revisions from complete new derived content whose canonical source record remains old. Both protected policies refuse service after failure; the first default retry restores all 80 fault runs through native full-rebuild fallback. In Hindsight, a shared PostgreSQL snapshot removes the mismatch in ten controlled pairs, with recall p95 rising from 19 ms to 29 ms. These results identify the read boundary needed for coherent evidence and the publication and recovery conditions governing service after an update fails.
\end{abstract}
\input{sections/01_introduction.tex}
\input{sections/02_evidence_and_service_criteria.tex}

\input{sections/03_read_boundaries_and_protection.tex}

\input{sections/04_evaluation_design.tex}

\input{sections/05_results.tex}

\input{sections/06_service_contracts_and_scope.tex}

\input{sections/07_related_work.tex}

\input{sections/08_conclusion.tex}

\input{sections/09_ai_use_statement.tex}

\input{references.tex}

\end{document}

%% file: sections/01_introduction.tex
\section{Introduction}
A memory service may return an observation with its supporting source. In one Hindsight execution \cite{r1}, the observation names Seattle, while resolving its parent identifier returns source text naming a different city. In a Cognee execution \cite{r2}, an old Seattle/Company X passage appears beside a new San Francisco/Company Y graph assertion. The identifiers link the records, but their contents come from different source revisions. Such a return gives a caller contradictory evidence for citation or audit.

Database isolation addresses incompatible reads \cite{r3}, \cite{r4}, yet the application determines which reads share its protection. A transaction on one connection does not cover enrichment on another; a lock released at request entry leaves later graph reads exposed. Protection must span the recall's first read through materialization of every lazy result. An update failure creates a further boundary: native work can terminate with derived content and its canonical source record in different publication states.

We protect complete recalls without changing the service's returned evidence. Hindsight connections share one PostgreSQL snapshot. For Cognee, we compare unprotected recall, a repeated recall (Double read), a read/write lock (Gate), and epoch validation (Epoch). Gate excludes writes through materialization; Epoch validates the completed read against a healthy generation. Both preserve an explicit failed state until a successful recovery reopens service.

The evaluation separates compatibility, completeness, freshness, waiting, and failure service. Under default-update load, Gate and Epoch deliver complete, compatible evidence for every scheduled recall, while unprotected recall conflicts in every tested workload run. Their reader and writer delays differ. After incremental failure, native reads can expose either conflicting revisions or complete new content ahead of canonical publication. The protected policies refuse these reads; a default retry recovers through the product's full-rebuild fallback. A shared snapshot addresses the corresponding cross-connection attribution problem in Hindsight.

The paper contributes:

\begin{itemize}
\item A characterization of attribution failures across PostgreSQL connections and of transient and failed-state publication boundaries across Cognee stores.
\item Implementations of recall-scoped snapshots, whole-call read/write protection, and bounded epoch validation, with update-completion and failure-state handling.
\item A 600-run Cognee comparison and a separate 160-run incremental-failure study, alongside Hindsight measurements, that connect evidence guarantees to availability, reader/writer delay, and actual recovery paths.
\end{itemize}

%% file: sections/02_evidence_and_service_criteria.tex
\section{Evidence and Service Criteria}
\subsection{Source compatibility}
A native return is the response assembled by the memory service. Backend and worker receipts identify the versions behind its fields; they are used only to check the response, not to construct it. Let $E$ be a return and $E_{\mathrm{cur}} = \pi_{\mathrm{cur}}(E)$ its items assigned current-state roles. Items explicitly marked as historical are not assigned current-state roles.

History $H$ records operations, known completion order, and source/parent relationships supported by the captured receipts. Let $\mathcal{C}(H)$ contain the dependency-consistent source states allowed by those records. We define compatibility as

\begin{equation}
C_1(E,H) \equiv \exists c \in \mathcal{C}(H)\;\forall e \in \pi_{\mathrm{cur}}(E):\operatorname{explains}(c,e). 
\end{equation}
For a raw item, $\operatorname{explains}(c,e)$ requires its bound source revision to agree with $c$. For a derived item, its recorded parent revision and publication visibility must also agree. For example, binding both Seattle and San Francisco as the current city of one source violates this contract. We evaluate $C_1$ on nonempty evidence with resolved bindings; other returns are marked \textnormal{\texttt{UNKNOWN}}. Cognee's checker validates local content relationships within this source scope; it does not reconstruct an atomic commit history across stores. For failed updates, we separately compare the delivered source basis with the observed canonical Data revision. A locally compatible and complete return can differ from that canonical record. This additional observation leaves the compatibility and usable-by-deadline definitions unchanged.

\subsection{Five complementary criteria}
Table I separates the questions asked of a repair. Cognee's completeness contract requires six roles: passage, summary, city and employer graph assertions, and city and employer fact text. A request is usable by deadline when it returns on time, has compatible source bindings, and contains all six. A response missing a role may still answer some questions; this study measures the representation contract, not answer quality.

\begin{table}[t]
\caption{Criteria used to evaluate a memory-service repair.}
\label{tab:I}
\centering
\footnotesize
\setlength{\tabcolsep}{3pt}
\begin{tabular}{@{}>{\raggedright\arraybackslash}p{0.20\columnwidth}>{\raggedright\arraybackslash}p{0.34\columnwidth}>{\raggedright\arraybackslash}p{0.40\columnwidth}@{}}
\toprule
Criterion & Question & Measurement \\
\midrule
Compatibility & Do the source revisions agree? & $C_1$ on bound evidence \\
Completeness & Are the required roles present? & Six role-presence flags \\
Freshness & How old is the source basis? & Lag against observed successful-update heads \\
Waiting & Where is time spent? & Queue, gate, read, and update durations \\
Failure service & What follows an update failure? & Partial delivery, refusal, and recovery probes \\
\bottomrule
\end{tabular}
\end{table}
Freshness measures lag against the successful-update heads observed at dispatch and at result materialization. Content can become visible before the update is acknowledged, so negative lag denotes that case. Hindsight's complete-and-current criterion, $C_2$, combines compatibility, its specified representation coverage, and zero observed source lag.

Service rates count every scheduled request, including errors, queue rejections, timeouts, refusals, and exhausted retries. Delivered-conflict rates instead use only classifiable delivered responses and are undefined when nothing is delivered. Empty responses are \textnormal{\texttt{UNKNOWN}} and are included in partial-return totals.

%% file: sections/03_read_boundaries_and_protection.tex
\section{Read Boundaries and Protection}
\subsection{Hindsight: one snapshot across connections}
Hindsight 0.10.2 retrieves observations before a later query resolves their parent identifiers to source text. A source edit between those stages can change the supporting content. On the tested edit path, the edit plan carries the existing \textnormal{\texttt{mentioned\_at}}, and the PostgreSQL update leaves that column unchanged. Before/after responses confirm that behavior. Thus the equal mention times in the example cannot order its source revisions.

Our wrapper binds all participating connections to one PostgreSQL view. A read-only Repeatable Read transaction exports a snapshot; each participating read-only transaction imports it before its first query. The exporter stays open through recall completion. PostgreSQL supplies snapshot export/import \cite{r5}, \cite{r6}, also used for parallel \textnormal{\texttt{pg\_dump}} \cite{r7}. The wrapper preserves retrieval SQL and parallel acquisitions. A curation check confirms the tested update path: an established snapshot sees the old source and observation, a fresh post-edit read sees the new source without the obsolete observation, and a settled read sees the matching new observation.

\subsection{Cognee: incremental publication and full rebuild}
Cognee 1.6.2 combines SQLite, LanceDB, and Ladybug in the tested configuration. One process owns the complete set of stores, which we call the scope. With answer generation disabled, hybrid search returns context, evidence objects, and references. The Python update API defaults \textnormal{\texttt{chunk\_level\_diff}} to \textnormal{\texttt{True}}; default requests omit the argument, whereas full-rebuild requests pass \textnormal{\texttt{False}}. Cognee also falls back to a full rebuild when an incremental update is not possible; branch logs distinguish incremental execution from fallback.

The publication sequences expose different intermediate states. Full rebuild withdraws representations before re-ingestion and reconstruction finish. Incremental update stages source content, writes new graph/vector content, withdraws replaced chunks, and publishes the canonical Data record. These are separate steps. A hybrid recall may therefore encounter old and new representations during publication, as the example in Fig. 1 shows.

\begin{figure*}[t]
\centering
\includegraphics[width=\textwidth]{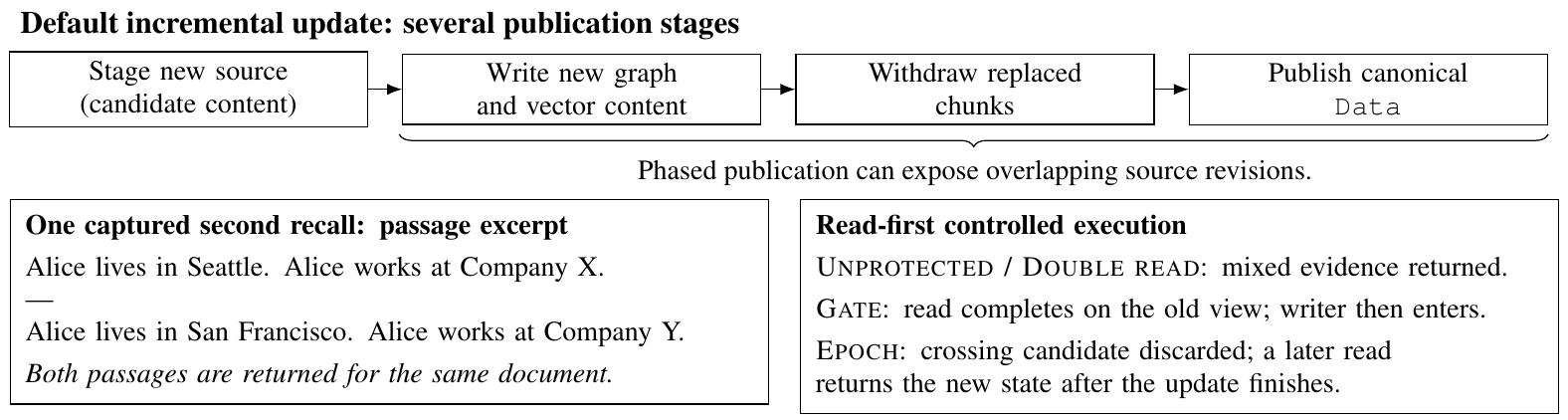}
\caption{Cognee's default update publishes through several stages. The arrows show operation order, not elapsed time. The excerpt is from the second recall in one controlled read-first execution; both passages are in the same returned context. The read-first and write-first schedules sample different stages, rather than delimiting a universal conflict window. Gate and Epoch descriptions refer to the same read-first ordering. The separate failure study injects P1 between new-content writing and chunk withdrawal, and P2 between withdrawal and canonical publication.}
\label{fig:1}
\end{figure*}
For chunks lacking \textnormal{\texttt{source\_content\_hash}}, the checker uses text fingerprints, document identity, acknowledged creation, and parent/source references. Unresolved bindings receive \textnormal{\texttt{UNKNOWN}}.

\subsection{Four read policies}
Table II summarizes the alternatives. UNPROTECTED is the service's ordinary recall; the original logs call it \textnormal{\texttt{NATIVE}}. DOUBLE READ repeats the whole recall and delivers the second result, without a checker, backoff, or fallback. All four policies use the same first-in, first-out (FIFO) writer queue.

\begin{table}[t]
\caption{Read policies. All preserve the service's evidence. Costs include every attempted read.}
\label{tab:II}
\centering
\footnotesize
\setlength{\tabcolsep}{3pt}
\begin{tabular}{@{}>{\raggedright\arraybackslash}p{0.20\columnwidth}>{\raggedright\arraybackslash}p{0.40\columnwidth}>{\raggedright\arraybackslash}p{0.34\columnwidth}@{}}
\toprule
Policy & Read during an update & Added work or waiting \\
\midrule
Unprotected & Collects visible evidence & One call \\
Double read & Runs the whole recall twice; delivers the second & Two calls if both are attempted \\
Gate & Existing readers finish; new readers wait & Reader/writer exclusion; readers share access \\
Epoch & Waits for a stable generation; validates after reading & Retries a crossed update; writer does not wait for read body \\
\bottomrule
\end{tabular}
\end{table}
GATE uses a writer-preference read/write lock. A reader holds permission through the awaited search and materialization: fully reading any lazy results into the response. A writer holds exclusive permission through the entire update, status/error checks, and relevant outstanding-task checks. Readers can proceed together; a queued writer prevents later readers from entering first.

EPOCH adapts the sequence-counter idea \cite{r8}. A short metadata lock protects generation, health, and operation identity. Before mutation, the writer advances to a higher odd generation marked \textnormal{\texttt{WRITING}}, then releases the metadata lock. Only the matching successful operation advances to an even \textnormal{\texttt{READY}} generation. A reader records a \textnormal{\texttt{READY}} generation, collects and materializes the response, and checks again. It accepts only the same healthy even generation. If an update crosses the read, the reader discards the candidate, including any error it carried; a native error under a stable generation is delivered. At most eight attempts fit within the original request deadline. Waiting consumes deadline time, and Epoch never falls back to Gate.

Epoch permits native reads and writes to overlap. The adapter retains Ladybug's lifecycle guards, which serialize submission against close and drain submitted work. Those guards cover the exercised access path; generation validation does not replace resource-lifetime protection. A fully materialized response remains fixed after validation even if a new update begins before delivery.

\subsection{Failed updates leave the scope \textnormal{\texttt{DIRTY}}}
Both protected policies keep the scope \textnormal{\texttt{DIRTY}} after an admitted mutation fails or its completion is uncertain. Releasing synchronization resources, cancelling an operation, or receiving a late success leaves that state intact. Recalls then receive an explicit refusal. A successful recovery in a newer generation restores \textnormal{\texttt{READY}}; admission depends on update status and operation identity, not the offline content checker. Rebuilding the data and reopening the read path are distinct steps.

%% file: sections/04_evaluation_design.tex
\section{Evaluation Design}
\subsection{Systems, workloads, and schedules}
The main Cognee evaluation comprises 600 fresh-store runs, with a separate 160-run incremental-failure study (Table III). Each uses one changing source with deterministic extraction and eight-dimensional embeddings. Each update replaces one chunk. The host runs Windows 11 with 16 logical CPUs and 31.2 GiB RAM, Python 3.12.14, LanceDB 0.39.0, Ladybug 0.19.0, and aiosqlite 0.22.1. Benchmark runs execute one at a time, with large offline analysis deferred until timing finishes.

\begin{table}[t]
\caption{Cognee run design. Each policy/update condition starts from an isolated copy of the same source history.}
\label{tab:III}
\centering
\footnotesize
\setlength{\tabcolsep}{3pt}
\begin{tabular}{@{}>{\raggedright\arraybackslash}p{0.23\columnwidth}>{\raggedright\arraybackslash}p{0.64\columnwidth}r@{}}
\toprule
Experiment & Paired variation & Runs \\
\midrule
Controlled & 20 histories \ensuremath{\times} 2 entrance orders \ensuremath{\times} 2 update modes \ensuremath{\times} 4 policies & 320 \\
Open arrival & 10 seeds \ensuremath{\times} 4 policies \ensuremath{\times} [read-only + 2 write profiles (spaced, burst) \ensuremath{\times} 2 modes] & 200 \\
Full-rebuild failure (main study) & 10 histories \ensuremath{\times} 4 policies \ensuremath{\times} [fault/control]; full rebuild & 80 \\
Incremental failure (separate study) & 10 histories \ensuremath{\times} 2 sites \ensuremath{\times} [fault/control] \ensuremath{\times} 4 policies & 160 \\
\bottomrule
\end{tabular}
\end{table}
Controlled read-first (AR) executions pause after the old passage is retrieved and before graph-neighborhood collection. Write-first (AW) executions pause after an acknowledged mutation: nonempty vector deletion for full rebuild, or the first graph/vector mutation in the incremental path. The scheduler releases operations at the publication or waiting event appropriate to each policy. Epoch's invalidation or stable-generation wait can release a paused writer, avoiding a circular wait for final read completion. A 90-second guard bounds these controlled holds.

Open-arrival runs submit approximately 20 recalls/s with 10\% interval jitter over 60 seconds. Policies share arrival lists, four active recall slots, 64 pending slots, one FIFO writer, and ten warmups. Two write profiles differ only in update timing: the spaced profile issues three updates about 21 seconds apart (at 6, 27, and 48 seconds), and the burst profile issues four updates within 0.3 seconds (at 21, 21.1, 21.2, and 21.3 seconds). The read-only workload is run once per policy/seed. Each of the five conditions has 12,010 scheduled recalls per policy across ten seeds.

Recall and update deadlines are 10 and 30 seconds after scheduled arrival, with a 165-second drain bound. Full queues reject immediately; active slots stay occupied until native termination, including after timeout. A monotonic nanosecond clock drives a rechecking wait loop: no request was dispatched early. Actual lateness, scheduled-arrival latency, and enqueue-based latency are all recorded; deadlines are never shifted to excuse a late request.

Failure runs inject one exception after the first completed nonempty vector deletion in a full rebuild. Five recalls probe the state at 0, 1, 2, 4, and 8 seconds after failure. A same-input rebuild starts at 15 seconds, followed by five recovery probes. Controls repeat the schedule without injection. The 15-second delay is an experimental trigger, separate from rebuild duration.

The separate incremental-failure study uses fresh stores and the same four policies and source change. Default requests omit \textnormal{\texttt{chunk\_level\_diff}}. At P1, a one-shot ordinary exception follows acknowledged new graph/vector writes and precedes withdrawal of the nonempty old-chunk set. At P2, it follows that withdrawal and precedes canonical Data publication (Fig. 1). Native error handling and fallback remain active; controls record the same sites without throwing. Each site, condition, and policy has ten histories, giving 160 runs.

Each run observes three stages: after the target update terminates (S0), after a same-input default retry (S1), and after a subsequent explicit full rebuild (S2). Five probes arrive at 0, 1, 2, 4, and 8 seconds after native termination, with a 10-second read deadline. A read-only storage checkpoint follows each probe group and records canonical and derived contents; it is an observer, not a protected service return. R1 and R2 each start at least 15 seconds after the preceding update terminates and after its probes and checkpoint finish. Every run follows this sequence, including controls. The 30-second update deadline and 120-second drain bound preserve task ownership. These sequential interventions measure actual recovery paths; they are not independently initialized recovery arms.

\subsection{Hindsight measurements}
The archived image/report and configuration identify Hindsight's PostgreSQL 17.11 and its 2--8 connection pool. Ten controlled pairs use four schedules in counts 3/3/2/2: ordinary update, prior supersession, held target derivation, and an intermediate update before held derivation. Twenty natural workloads per condition each complete fifteen writes and at least sixty recalls and run for at least fifteen seconds, within a ninety-second ceiling. These recall streams wait for the previous capture to finish.

Snapshot cost is measured on five no-write pairs, each with five warmups and one hundred measured recalls after consolidation. Native-read time includes the recall and its connection behavior; total-request time also includes exporter handling, mapping, and durable logging. The predeclared cost screen permits a total p95 ratio of at most 1.25 and a client-rate ratio of at least 0.80. These costs are auxiliary to the snapshot correctness result.

\subsection{Statistical summaries}
We compute each workload's service rates and 95th-percentile latency (p95) before pairing seeds. Confidence intervals (CIs) are descriptive 95\% percentile intervals from 10,000 paired cluster resamples: twenty histories within each controlled condition, ten whole runs per workload condition, and ten failure histories with probes nested. Means of run quantiles, pooled counts, and medians are labeled separately. An update-run p95 summarizes only three spaced or four burst writes. Plot bars show medians and interquartile ranges (IQRs), not CIs. Each study froze its execution and analysis plan before its first formal run; all planned policy and mode contrasts are reported. Incremental-failure rates pair ten histories per site and condition, with the five stage probes nested within each history. Its intervals use 10,000 paired-history resamples (seed 2026111302). The supplemental run is reported separately from the original performance comparison.

%% file: sections/05_results.tex
\section{Results}
All 600 main-study runs completed. Every one of the 440 default updates executed incrementally without fallback. The full-rebuild group had 560 successful rebuilds and forty injected failures. Each successful update published its target source content.

\subsection{A second recall can still cross a publication window}
Double read still returns conflicting evidence during default incremental publication. In all twenty default read-first histories, both the unprotected response and the delivered second response conflict (Table IV). The example in Fig. 1 combines old and new passages, mixed graph assertions, and old facts. Repeating the call therefore did not, by itself, wait out this controlled publication window.

\begin{table}[t]
\caption{Controlled recalls: twenty histories per cell. U = usable, X = directly conflicting, P = partial. AR and AW denote read-first and write-first; Full and Default denote full rebuilds and default incremental updates. All primary calls return; schedule counts are not natural frequencies.}
\label{tab:IV}
\centering
\footnotesize
\setlength{\tabcolsep}{3pt}
\begin{tabular*}{\columnwidth}{@{\extracolsep{\fill}}lcccc@{}}
\toprule
Policy & \shortstack{Full\\AR} & \shortstack{Full\\AW} & \shortstack{Default\\AR} & \shortstack{Default\\AW} \\
\midrule
Unprotected & 20 X & 20 P & 20 X & 20 U \\
Double read & 20 P & 20 P & 20 X & 20 U \\
Gate & 20 U & 20 U & 20 U & 20 U \\
Epoch & 20 U & 20 U & 20 U & 20 U \\
\bottomrule
\end{tabular*}
\end{table}
Both protected policies return complete evidence. Gate completes the acquired read on the old state before admitting the writer. Epoch discards a crossing candidate, waits for a stable generation, and accepts a new read. Across both update modes, its forty read-first executions discard forty conflicting candidates. Under full rebuild, Double read's second call becomes compatible but lacks fact roles. In the default write-first schedule, unprotected readers can obtain a complete old response at the chosen first-mutation pause. This observation describes that pause, not every intermediate stage of incremental publication.

\subsection{Default updates trade missing representations for mixed versions}
Under open-arrival load, default updates preserve all six roles but can return contradictory versions. Unprotected recall conflicts in every one of its twenty default-update runs: 199 of 12,010 scheduled recalls with spaced writes (1.66\%) and 284 of 12,010 with bursts (2.36\%). Double read also returns conflicts. Full rebuilds produce partial evidence and caller errors, with no directly conflicting delivery in these workloads. Table V reports these outcomes together with usable service and rejections.

In paired comparisons with full rebuilds, default updates raise the unprotected conflict fraction by 1.66 percentage points (pp), 95\% CI [1.57, 1.75], for spaced writes and by 2.36 pp [2.22, 2.56] for bursts. Incremental processing avoids the observed deletion gaps, while phased publication still permits mixed-version returns.

Gate and Epoch each provide complete, compatible responses for all 24,020 scheduled default-update recalls. With full-rebuild bursts, Gate rejects 64 requests in one seed and Epoch serves all requests. Their paired usable-fraction difference is 0.53 pp, 95\% CI [0, 1.60]. Unprotected, Gate, and Epoch also serve every read-only request; Double read's 2,929 read-only rejections show the cost of unconditional extra work at this offered load.

\begin{table*}[t]
\caption{Open-arrival outcomes and work. Every row contains ten runs and 12,010 scheduled recalls; Full and Default denote full rebuilds and default incremental updates. Usable, conflict, partial, error, and rejection counts exhaust those requests in this experiment. UNKNOWN is included in partial. Attempts and discards count native calls, not delivered responses.}
\label{tab:V}
\centering
\footnotesize
\setlength{\tabcolsep}{3pt}
\begin{tabular*}{\textwidth}{@{\extracolsep{\fill}}llrrrrrrr@{}}
\toprule
Condition & Policy & Usable & Conflict & Partial & Error & Rejected & Calls & Discarded \\
\midrule
Read only & Unprotected & 12,010 & 0 & 0 & 0 & 0 & 12,010 & 0 \\
Read only & Double read & 9,081 & 0 & 0 & 0 & 2,929 & 18,162 & 0 \\
Read only & Gate & 12,010 & 0 & 0 & 0 & 0 & 12,010 & 0 \\
Read only & Epoch & 12,010 & 0 & 0 & 0 & 0 & 12,010 & 0 \\
\midrule
Full, spaced & Unprotected & 11,685 & 0 & 144 & 181 & 0 & 12,010 & 0 \\
Full, spaced & Double read & 8,724 & 0 & 116 & 410 & 2,760 & 18,500 & 0 \\
Full, spaced & Gate & 12,010 & 0 & 0 & 0 & 0 & 12,010 & 0 \\
Full, spaced & Epoch & 12,010 & 0 & 0 & 0 & 0 & 12,063 & 53 \\
\midrule
Full, burst & Unprotected & 11,541 & 0 & 191 & 278 & 0 & 12,010 & 0 \\
Full, burst & Double read & 8,558 & 0 & 170 & 536 & 2,746 & 18,528 & 0 \\
Full, burst & Gate & 11,946 & 0 & 0 & 0 & 64 & 11,946 & 0 \\
Full, burst & Epoch & 12,010 & 0 & 0 & 0 & 0 & 12,147 & 137 \\
\midrule
Default, spaced & Unprotected & 11,759 & 199 & 0 & 0 & 52 & 11,958 & 0 \\
Default, spaced & Double read & 8,746 & 126 & 0 & 0 & 3,138 & 17,744 & 0 \\
Default, spaced & Gate & 12,010 & 0 & 0 & 0 & 0 & 12,010 & 0 \\
Default, spaced & Epoch & 12,010 & 0 & 0 & 0 & 0 & 12,063 & 53 \\
\midrule
Default, burst & Unprotected & 11,726 & 284 & 0 & 0 & 0 & 12,010 & 0 \\
Default, burst & Double read & 8,927 & 166 & 0 & 0 & 2,917 & 18,186 & 0 \\
Default, burst & Gate & 12,010 & 0 & 0 & 0 & 0 & 12,010 & 0 \\
Default, burst & Epoch & 12,010 & 0 & 0 & 0 & 0 & 12,149 & 139 \\
\bottomrule
\end{tabular*}
\end{table*}
\begin{figure*}[t]
\centering
\includegraphics[width=\textwidth]{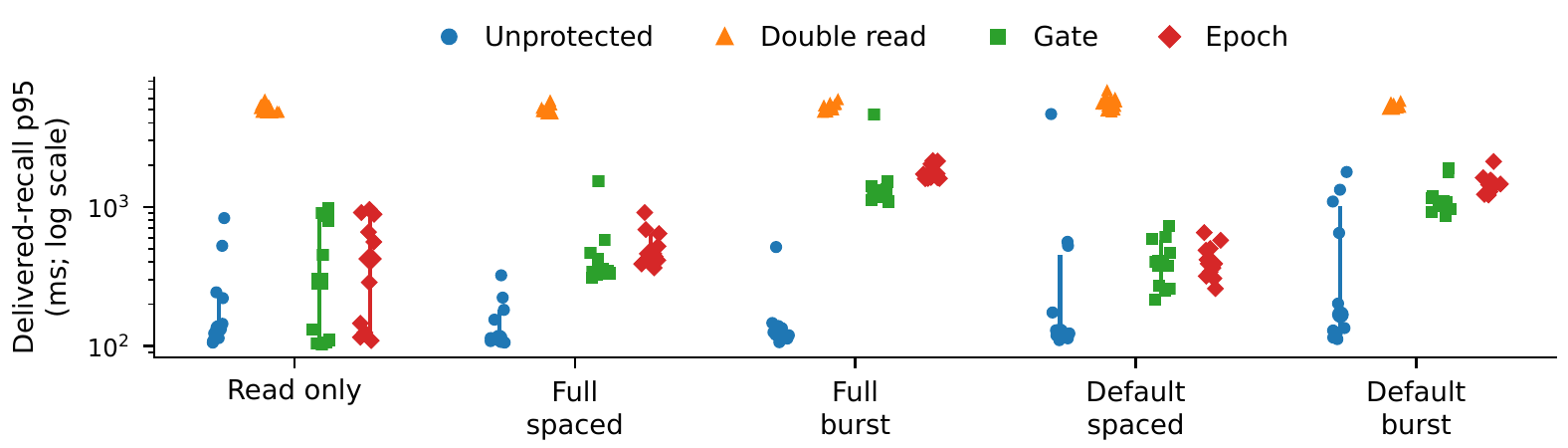}
\caption{Delivered-recall p95 across all 200 open-arrival runs. Jittered points show individual seeds; centered markers and vertical whiskers show the median and IQR. Time starts at scheduled arrival. Queue-rejected requests are excluded from this latency measure but remain in Table V. The logarithmic axis shows both normal service and double-read queue pressure.}
\label{fig:2}
\end{figure*}
\subsection{Epoch shifts cost between readers, writers, and repeated work}
Epoch exchanges shorter update tails for longer recall tails under default bursts. Mean run recall-p95 is 1,475 ms for Epoch and 1,185 ms for Gate: a paired increase of 290 ms, 95\% CI [48, 562]. Mean run update-p95 is lower by 109 ms [\ensuremath{-}168, \ensuremath{-}64]. Epoch lets the writer proceed without waiting for an active read body, but readers may wait for stability and repeat an invalidated call. Thus the shorter update tail comes with a longer recall tail; quantile differences do not identify its individual causes.

For spaced updates, the 95\% CI of the recall-tail difference between Epoch and Gate includes zero under both update modes. With full-rebuild bursts, the paired difference is 188 ms, 95\% CI [\ensuremath{-}445, 595]; in this condition, Gate also rejects requests. Fig. 2 shows the variation across seeds rather than a single aggregate latency.

Epoch discards 382 candidates under open-arrival load: 53 and 137 for full-rebuild spaced and burst updates, and 53 and 139 for default spaced and burst updates. All were compatible and error-free in the offline checks; they were discarded because publication crossed their reads. This small discard count does not measure prevented errors. The forty discarded controlled conflicts form a separate set. No stable native error or retry-limit exhaustion occurred.

Queueing is part of the protection cost, but also reflects host congestion. In the Gate run with rejections under full-rebuild bursts, the first rejection occurs at 13.809 s, before the first scheduled update at 21 s. Four native recalls occupy the active slots and 64 requests are pending. Later, a selected request scheduled for 18.885 s waits until 23.174 s for a slot, then until 23.745 s for the fourth update to finish, and completes native retrieval at 23.898 s. Its 4.289 s external-queue wait and 0.571 s gate wait are distinct intervals. Gate-blocked readers can occupy active slots and propagate waiting into that queue; this trace also shows why not all congestion can be assigned to a writer lock.

All 240,200 workload requests are accounted for, including 14,606 queue rejections, with zero early dispatches, negative elapsed durations, or read-deadline outcomes. Among returned responses with interpretable source bindings, lag is known for 223,245 and unresolved for 944. No known lag is positive; 231 responses contain content visible before the corresponding update was acknowledged. These observations are relative to successful-update heads. A mixed or unbound response does not acquire a freshness guarantee from this summary.

\subsection{Publication state determines failed-service behavior and recovery}
In the main full-rebuild failure experiment, each unprotected policy delivers fifty partial probes after deletion, and each protected policy refuses fifty while \textnormal{\texttt{DIRTY}}. All four have zero usable responses at this stage. After the same-input rebuild, each delivers fifty usable probes; no-fault controls remain usable before and after their repeat rebuild.

The incremental-failure study completes all 160 runs, 2,400 probes, and 480 study-update calls. Every target request enters the incremental route. All 80 fault targets reach their designated site and fail through native error handling; all 80 controls finish incrementally. After P1, the checkpoints retain both old and new graph/vector content while canonical Data remains old. After P2, the derived content is new and canonical Data is still old. These states persist through the observed post-termination probes and checkpoint.

\begin{table}[t]
\caption{Service after incremental failure, before retry. Each policy/site cell contains ten histories and fifty probes. U = locally usable, X = directly conflicting, R = DIRTY refusal, M = delivered basis differs from the observed canonical Data. U and M can overlap; refusal has no delivered basis.}
\label{tab:VI}
\centering
\footnotesize
\setlength{\tabcolsep}{3pt}
\begin{tabular*}{\columnwidth}{@{\extracolsep{\fill}}l l rrrr@{}}
\toprule
Site & Policy & U & X & R & M \\
\midrule
P1 & Unprotected & 0 & 50 & 0 & 50 \\
P1 & Double read & 0 & 50 & 0 & 50 \\
P1 & Gate & 0 & 0 & 50 & 0 \\
P1 & Epoch & 0 & 0 & 50 & 0 \\
P2 & Unprotected & 50 & 0 & 0 & 50 \\
P2 & Double read & 50 & 0 & 0 & 50 \\
P2 & Gate & 0 & 0 & 50 & 0 \\
P2 & Epoch & 0 & 0 & 50 & 0 \\
\bottomrule
\end{tabular*}
\end{table}
Unprotected and Double read each deliver fifty conflicting P1 probes. At P2, each delivers fifty locally complete, compatible probes whose source basis differs from canonical Data (Table VI). Gate and Epoch refuse every fault-stage probe. At P1, refusal prevents conflicting delivery; at P2, it withholds coherent new derived content pending successful recovery. Refusal preserves publication-aware admission at the cost of availability while the scope remains \textnormal{\texttt{DIRTY}}.

The first default retry restores canonical v1 and all five usable probes in every fault run. All 80 retries take the native full-rebuild fallback with reason \textnormal{\texttt{chunks\_not\_tiling}}; the 80 control retries return \textnormal{\texttt{unchanged}}. Thus the request remains default while the successful recovery route is a rebuild. Gate and Epoch reopen through the update-status and operation-identity checks. The prescribed R2 explicitly rebuilds all 160 stores, after which all S2 probes remain usable. It adds no observed content benefit after successful R1. The fixed 15-second trigger, native recovery body, and observer/safety waits remain separate intervals.

This comparison identifies the service contract enforced by the failed-state guard: callers receive an explicit refusal until a successful operation restores publication and service. The underlying content state can be mixed or locally coherent, so compatibility, canonical agreement, and availability must be reported separately.

\subsection{Hindsight snapshots repair the cross-connection attribution}
Sharing one PostgreSQL snapshot removes the Hindsight mismatch in all ten controlled pairs: ten unprotected recalls conflict, and none of their snapshot counterparts does. The protected response uses an older compatible view, while a later settled read meets the complete-and-current criterion. Natural workloads yield too few events to compare frequency: one conflict in 2,551 unprotected recalls and none in 2,241 snapshot recalls. The workload-paired mean difference is \ensuremath{-}0.397 events per thousand recalls, 95\% CI [\ensuremath{-}1.190, 0].

Across five no-write pairs, median native-read p95 rises from 19.09 to 28.97 ms (1.52-fold); total-request p95 rises from 78.51 to 94.27 ms. The median paired total-p95 ratio is 1.206, within the predeclared 1.25 limit, but the one-client recall-rate ratio is 0.790, below the 0.80 screen. The latency condition passes and the client-rate condition does not; the joint screen is therefore unmet. The snapshot supplies a consistent view at the measured request cost. These rates measure one-client completion, not concurrent capacity.

These counts rely on direct source/parent evidence. Fifty-four unprotected and 128 snapshot labels inferred only from publication order are excluded, because discovering an observation after an edit does not establish when it was published. Their unresolved status remains recorded.

%% file: sections/06_service_contracts_and_scope.tex
\section{Service Contracts and Deployment Scope}
Table VII summarizes what each criterion reveals. Snapshots retain a database view, Gate prevents overlapping operations, and Epoch validates a completed read.

\begin{table*}[t]
\caption{Findings across five criteria. Hindsight and Cognee use different measurement protocols; the table compares protection choices rather than ranking their speed.}
\label{tab:VII}
\centering
\footnotesize
\setlength{\tabcolsep}{3pt}
\begin{tabular}{@{}>{\raggedright\arraybackslash}p{0.13\textwidth}>{\raggedright\arraybackslash}p{0.33\textwidth}>{\raggedright\arraybackslash}p{0.50\textwidth}@{}}
\toprule
Criterion & Hindsight shared snapshot & Cognee comparison \\
\midrule
Compatibility & Controlled conflicts: 10/10 unprotected, 0/10 snapshot. & Default-update workloads: 199 spaced and 284 burst conflicts under Unprotected; zero delivered conflicts under Gate/Epoch. \\
Completeness & All 35 terminal checks per variant meet the complete/current target: ten controlled, twenty natural, five timing runs. & Unprotected full rebuild has 144 spaced and 191 burst partial returns; Gate/Epoch deliver no partial response. Gate also rejects 64 recalls under full-rebuild bursts. \\
Freshness & A snapshot can preserve the pre-update view. & No positive lag among 223,245 interpretable returns; 944 unresolved. New content can precede acknowledgement. \\
Waiting/cost & Native-read p95 rises 1.52-fold; client-rate ratio 0.790 against a 0.80 screen. & Default burst, Epoch relative to Gate: recall p95 +290 ms, update p95 \ensuremath{-}109 ms; 139 discarded attempts. \\
Failure service & Failed-update recovery was not evaluated. & Full-rebuild failure leaves partial evidence. Incremental failure leaves mixed revisions or new derived content ahead of canonical publication. Gate/Epoch refuse; rebuilding restores the observed service. \\
\bottomrule
\end{tabular}
\end{table*}
\paragraph{Publication-aware service after failure}
A failed update can leave partial evidence, mixed revisions, or locally coherent new content ahead of canonical publication. Gate and Epoch enforce a service contract that requires a successful recovery before reopening. A \textnormal{\texttt{DEGRADED}} interface could expose available content together with missing roles or publication status if callers can interpret those distinctions. Automatically queuing a rebuild would also require retry and resource-use rules. The tested guard preserves \textnormal{\texttt{DIRTY}} until an explicit successful recovery; without that trigger it can persist indefinitely.

\paragraph{Build then swap}
Building a new generation while serving the old one is another way to protect publication. A hybrid recall would need to retain one generation of all database handles and caches; publication and old-generation reclamation would need coordinated lifetimes. This design adds storage, recovery, and resource-reclamation work. Atomic publication would also need to be paired with a consistent read view across the recall's stages.

\paragraph{Multiple worker processes}
Our scope has one owner process. Extending either policy to several workers requires shared generation and health state, participation by every writer and background publisher, and safe backend-resource lifetimes. A shared lock alone would not supply all three. Epoch additionally depends on the stores tolerating the read/write overlap it validates.

\paragraph{Experimental scope}
The experiments use fixed versions, deterministic extraction and embeddings, and a single changing source. Controlled schedules establish possible executions; their repetitions are not independent semantic tasks or incident-rate samples. Fault coverage comprises the full-rebuild deletion site and two incremental publication sites, using ordinary exceptions rather than process crashes. Six-role coverage is a service contract rather than downstream answer quality. Neither policy repairs pre-existing inconsistency or guarantees freshness until later consumption. The CIs describe the sampled histories and workloads; update-run tail estimates use three or four writes. The desktop host had 1.71 GiB available memory in a pre-run snapshot, so hardware capacity should not be read as a dedicated-memory reservation; the cause of an early congestion episode is unmeasured. Hindsight's exact contemporaneous hardware and Python/asyncpg microversions are unavailable, which limits reproduction of its auxiliary cost measurements. We did not test its serial single-connection Repeatable Read alternative, exhaustive cancellation/import failure, multi-owner safety, or downstream model behavior.

\paragraph{Data availability}
Code and data are available at \url{https://github.com/hyauu/read-boundaries-artifact}.

%% file: sections/07_related_work.tex
\section{Related Work}
Read skew and snapshot isolation have established database definitions \cite{r3}. RAMP provides atomic visibility across partitions through transaction metadata and versioned retrieval \cite{r4}. PostgreSQL synchronizes views across connections \cite{r5}--\cite{r7}; sequence counters detect reads that cross a write while requiring separate resource-lifetime protection \cite{r8}. We locate the operation boundaries at which these existing ideas apply in working memory services and measure the tradeoffs of protecting them. Elle's reconstruction of dependencies from observations \cite{r9} motivates separating direct source bindings from uncertain publication order.

Agent-memory work addresses related guarantees at different layers. MemTX v2 defines middleware with belief admission, action gating, and read-visibility levels: committed-read excludes staged records, and dependency-stability checks can reject records with ancestors pending revocation \cite{r10}. MemTxn combines source-supported update admission, temporal version selection, and recovery of application-visible state \cite{r11}. Khan studies verified concurrency-anomaly detection and prevention under deterministic-generation semantics \cite{r12}. SemIso specifies resource bindings and compatibility for durable AI workflow continuation \cite{r13}. We examine evidence collected by existing Hindsight and Cognee read paths and measure how protecting those calls changes service behavior. End-to-end model performance of these frameworks is outside this comparison.

Zhang \cite{r14} examines conflicting outputs from security tools and uses an independent evidence check to verify candidate decisions. Chaughule et al. \cite{r15} trace decisions in an agentic cryptocurrency-monitoring pipeline to individual components through interventions on their outputs. These studies address evidence reliability and decision responsibility at the agent layer. The read boundaries studied here govern which source revisions a memory service returns together during updates and recovery.

%% file: sections/08_conclusion.tex
\section{Conclusion}
Whole-recall protection preserves the relationship between returned evidence and its source revision across the tested memory-service boundaries. Hindsight's shared snapshot gives participating connections one view. In Cognee, Gate and Epoch deliver complete, compatible evidence throughout the measured default-update workloads, with distinct reader and writer costs. Incremental failures expose a further separation: derived content can be coherent even when canonical publication has not completed. The failed-state guard makes that boundary explicit to callers, and the native rebuild path restores publication and service. Read consistency, publication agreement, and recovery admission together determine what a caller can rely on after an update.

%% file: sections/09_ai_use_statement.tex
\section*{AI Use Statement}
AI tools were used for implementation, analysis, and manuscript polishing.

%% file: references.tex
\balance